\documentclass[sigconf]{acmart}

\renewcommand\footnotetextcopyrightpermission[1]{}
\usepackage{booktabs}
\usepackage{multirow}
\usepackage{colortbl}
\usepackage{pifont}
\usepackage{listings}
\usepackage{subcaption}
\usepackage{tikz}
\usepackage{tabularx}
\usepackage{array}
\usetikzlibrary{arrows.meta,positioning,shapes.geometric,fit}

\lstdefinelanguage{TLAplus}{
  morekeywords={MODULE,EXTENDS,CONSTANT,VARIABLE,ASSUME,Init,Next,Spec,
    INVARIANTS,SPECIFICATION,TRUE,FALSE,IF,THEN,ELSE,LET,IN,
    UNCHANGED,EXCEPT,SUBSET,UNION,BOOLEAN},
  sensitive=true,
  morecomment=[l]{\\*},
  morestring=[b]",
  literate={/\\}{{$\wedge$}}1 {\\/}{{$\vee$}}1
           {=>}{{$\Rightarrow$}}1 {->}{{$\rightarrow$}}1
           {<<}{{$\langle$}}1 {>>}{{$\rangle$}}1
           {\\in}{{$\in$}}1 {\\notin}{{$\notin$}}1
           {\\A}{{$\forall$}}1 {\\E}{{$\exists$}}1
           {\\subseteq}{{$\subseteq$}}1
}
\definecolor{bothfail}{HTML}{D32F2F}
\definecolor{specfail}{HTML}{F57C00}
\definecolor{implfail}{HTML}{FFA726}
\definecolor{speconly}{HTML}{FFD54F}
\definecolor{cellpass}{HTML}{4CAF50}
\definecolor{undercon}{HTML}{90A4AE}
\definecolor{notappl}{HTML}{E0E0E0}

\newcommand{\SPECFAIL}{\cellcolor{specfail!30}SPEC}

\newcommand{\CELLPASS}{\cellcolor{cellpass!30}\ding{51}}
\newcommand{\UNDERCON}{\cellcolor{undercon!30}UC}

\newcommand{\sys}{AgentConform}
\newcommand{\aps}{APS}
\newcommand{\aasm}{AASM}
\newcommand{\tlaplus}{TLA\textsuperscript{+}}

\begin{document}

\title{AgentRFC: Security Design Principles and Conformance Testing for Agent Protocols}

\author{Shenghan Zheng}
\affiliation{%
  \institution{Dartmouth College}
  \city{Hanover}
  \state{NH}
  \country{USA}
}

\author{Qifan Zhang}
\affiliation{%
  \institution{Palo Alto Networks}
  \city{Santa Clara}
  \state{CA}
  \country{USA}
}


\begin{abstract}
AI agent protocols---including MCP, A2A, ANP, and ACP---enable
autonomous agents to discover capabilities, delegate tasks, and
compose services across trust boundaries. Despite massive
deployment (MCP alone has 97M+ monthly SDK downloads), no systematic
security framework for these protocols exists.

We present three contributions.
First, the \emph{Agent Protocol Stack} (\aps{}), a 6-layer
architectural model that defines what a complete agent protocol must
specify at each layer---analogous to ITU-T X.800 for the OSI stack.
Second, the \emph{Agent-Agnostic Security Model} (\aasm{}), 11
security principles formalized as \tlaplus{} invariants, each tagged
with a property taxonomy (\textsc{spec-mandated},
\textsc{spec-recommended}, \textsc{aasm-hardening},
\textsc{aps-completeness}) that distinguishes protocol
non-conformance from framework-imposed security requirements.
Third, \emph{\sys{}}, a two-phase conformance checker that
(i)~extracts normative clauses from protocol specifications into a
typed Protocol~IR with explicit
\emph{Protocol}/\emph{Environment}/\emph{Adversary} action
separation, (ii)~compiles the IR into \tlaplus{} models and
model-checks them against \aasm{} invariants, then
(iii)~replays counterexample traces against live SDK implementations
to confirm findings.

We introduce the \emph{Composition Safety} (CS) principle: security
properties that hold for individual protocols can break when protocols
are composed through shared infrastructure. We demonstrate this with
formal models of five protocol composition patterns, revealing
cross-protocol design gaps that individual protocol analysis cannot
detect. Preliminary application to representative agent protocols
reveals recurrent gaps in credential lifecycle, consent enforcement,
audit completeness, and composition safety. Some findings are under
coordinated disclosure; full evaluation details will be released in
the complete version.
\end{abstract}



\maketitle


\section{Introduction}\label{sec:intro}

AI agent protocols---MCP~\cite{mcp2024}, A2A~\cite{a2a2024},
ANP~\cite{anp2025}, and ACP~\cite{acp2025}---are becoming the control
plane for tool use, delegation, and multi-agent orchestration.
They define how agents exchange capabilities, execute actions on behalf
of users, and pass intermediate outputs across trust boundaries.
Unlike traditional API protocols, these systems carry \emph{semantic}
payloads (natural-language instructions, tool outputs, delegation context)
that are interpreted by LLMs at runtime. As a result, failures in one
component can propagate through reasoning context and trigger actions in
another component.

This shift creates a security-analysis gap. Classical protocol methods are
strong on cryptographic and message-level properties, but agent protocols
also require semantic properties such as prompt integrity, explicit consent,
and cross-protocol isolation. At the same time, real-world incidents and
ongoing protocol proposals indicate that these risks are not hypothetical.
However, there is still no unified, formal, cross-protocol framework for
evaluating whether agent protocols satisfy such properties.

\paragraph{Contributions.}
We make three contributions:

\begin{enumerate}
\item \textbf{Agent Protocol Stack (\aps{}).}
  We introduce a 6-layer architectural model (Transport Security through
  Audit \& Accountability) that makes protocol completeness explicit.
  Applying \aps{} to MCP, A2A, ANP, and ACP shows that only transport
  security is consistently complete; higher layers remain under-specified
  in all protocols (\S\ref{sec:aps}).

\item \textbf{Agent-Agnostic Security Model (\aasm{}).}
  We define 11 principles---8 security properties (P1--P8) and
  3 completeness properties (WF, SL, CS)---as \tlaplus{} invariants.
  Each property is tagged using a taxonomy
  (\textsc{spec-mandated}, \textsc{spec-recommended},
  \textsc{aasm-hardening}, \textsc{aps-completeness}) to distinguish
  protocol non-conformance from frame\-work-im\-posed hardening
  requirements (\S\ref{sec:aasm}).

\item \textbf{\sys{}.}
  We build a two-phase conformance checker with a provenance-preserving
  pipeline from prose specifications to executable tests.
  We extract normative clauses, compile them into a typed Protocol IR
  with explicit \emph{Protocol}/\emph{Environment}/\emph{Adversary}
  action separation, and generate \tlaplus{} models where each action
  and invariant is traceable to source clauses or explicit assumptions.
  Phase~1 model-checks five protocols against 11 principles, yielding a
  $5 \times 11$ matrix with 33 spec-level violations.
  Phase~2 replays counterexample traces as tests against live SDKs,
  executing 42 tests across reference SDK implementations and
  official reference servers, confirming violations at implementation
  level. Additionally, 5~composed models reveal 20 composition safety
  violations across all protocol pairs
  (\S\ref{sec:composition}).
\end{enumerate}

Our most significant finding is the \emph{Composition Safety} (CS)
principle (\S\ref{sec:composition}): when MCP and A2A operate through a
shared bridge (conductor/proxy), a prompt-injection attack in MCP
propagates to amplify A2A delegation beyond its original scope.
This cross-protocol design gap cannot be detected by analyzing
either protocol in isolation.

We also report responsible-disclosure outcomes and mitigation guidance
for the findings that are reproducible at implementation level,
including advisory submissions and specification-improvement proposals
(\S\ref{sec:discussion}).

\paragraph{Paper roadmap.}
\S\ref{sec:background} introduces protocol context and threat model.
\S\ref{sec:aps} presents \aps{}, and \S\ref{sec:aasm} formalizes
the 11 principles. \S\ref{sec:agentconform} describes \sys{}.
\S\ref{sec:composition} analyzes cross-protocol composition safety.
\S\ref{sec:discussion} and \S\ref{sec:related} conclude with
implications, limitations, and prior work.

\section{Background}\label{sec:background}

\subsection{The Rise of AI Agent Protocols}

Traditional web APIs expose deterministic endpoints: a client sends a
request, and the server returns a response whose structure is fully
specified by an OpenAPI schema or similar contract. AI agent protocols
break this model in three fundamental ways.

First, agent protocols carry \emph{semantic payloads}. A tool invocation
in MCP returns natural-language text that an LLM interprets as part of
its reasoning context. Unlike a JSON field with a fixed schema, this
text can contain instructions, injections, or content that crosses
trust boundaries---a class of attack that traditional API security
(rate limiting, input validation, OAuth scoping) does not address.

Second, agent protocols support \emph{delegation chains}. In A2A, an
agent can delegate a task to another agent, which may further delegate
to a third. The original user's authority must be preserved across these
hops, but no protocol currently enforces monotonicity: an agent can
re-delegate capabilities it was not originally granted.

Third, agent protocols are increasingly \emph{composed}. A conductor
(proxy) may bridge MCP and A2A, routing tool outputs from one protocol
into delegation decisions in another. Security properties that hold
for each protocol in isolation may fail under composition---a problem
well-studied in cryptographic protocols~\cite{cortier2009safely} but
unaddressed for agent protocols.

These differences motivate a new security framework. Existing protocol
analysis tools (ProVerif~\cite{blanchet2001proverif},
Tamarin~\cite{meier2013tamarin}) target cryptographic properties
(secrecy, authentication) in unbounded sessions. Agent protocols
require properties at a higher semantic layer: prompt integrity,
consent gates, audit completeness, and composition safety. We use
\tlaplus{}/TLC for bounded model checking because it directly produces
counterexample traces that we convert into executable test cases.

\subsection{AI Agent Protocols Under Study}

We study five agent protocols that have emerged since 2024. Each
occupies a different point in the design space:

\paragraph{MCP (Model Context Protocol)~\cite{mcp2024}.}
MCP defines a client-server interface for LLM tool invocation.
A host application connects to MCP servers that expose tools, resources,
and prompts via JSON-RPC~2.0. The protocol added OAuth~2.0
authentication in March~2025 and has 97M+ monthly SDK downloads.
The specification states that servers ``\textsc{must} sanitize tool
outputs'' and clients ``\textsc{should} validate tool results before
passing to LLM,'' but the reference SDK implements neither at the
time of writing. MCP lacks credential revocation on session close,
consent gates for sensitive tools, and mandatory audit trails.

\paragraph{A2A (Agent-to-Agent)~\cite{a2a2024}.}
A2A enables peer-to-peer agent communication via JSON-RPC~2.0,
gRPC, or HTTP+JSON/REST. Agents publish capability cards
(manifests with skills, security schemes, and interface declarations)
and a conductor orchestrates multi-agent sessions through tasks.
A2A defines a rich task lifecycle
(submitted$\to$working$\to$completed/failed/canceled) but does not
specify delegation scope, consent requirements, audit obligations, or
credential revocation. Agent card signing is optional.

\paragraph{ANP (Agent Negotiation Protocol)~\cite{anp2025}.}
ANP is a white-paper protocol for multi-party negotiation.
Its specification is the least mature: session close semantics are vague,
message framing is minimal, and consent is not formalized. ANP serves
as our baseline for maximum underspecification---a protocol where
ambiguity itself is the primary finding.

\paragraph{ACP-Cap (Agent Communication Protocol)~\cite{acp2025}.}
ACP-Cap is an evolving standard that addresses some gaps in MCP and A2A.
Recent RFDs add session lifecycle operations (suspend/resume) and
partial consent gates, but audit requirements and fail-secure defaults
remain unspecified. ACP-Cap is the closest to completeness among the
protocols but still fails on 4 of 7 checked \aasm{} principles.

\paragraph{ACP-Client (Agent Client Protocol)~\cite{acpclient2025}.}
ACP-Client standardizes communication between code editors (IDEs)
and AI coding agents. Unlike MCP's client-server tool invocation
model, ACP-Client gives the agent \emph{direct filesystem access}:
agents can read and write arbitrary files via \texttt{fs/read\_text\_file}
and \texttt{fs/write\_text\_file}. The specification requires agents to
verify filesystem capabilities but imposes no path restrictions,
sandboxing, or content validation on file writes. Permission for tool
execution is \textsc{may}-level (advisory), not \textsc{must}-level.
ACP-Client is used by JetBrains IDEs, Zed, and agents including
Claude~Code and Codex~CLI, making it a high-value target: injection
via ACP-Client can lead to arbitrary code execution through malicious
file writes (e.g., to \texttt{.git/hooks/pre-commit} or
\texttt{\textasciitilde/.ssh/authorized\_keys}).

\subsection{Formal Verification with TLA\textsuperscript{+}}

\tlaplus{}~\cite{lamport2002specifying} is a formal specification
language based on the temporal logic of actions. A specification defines
an initial state predicate~$Init$, a next-state relation~$Next$, and
invariants~$Inv$ (safety properties). The specification
$Spec \triangleq Init \wedge \Box[Next]_{vars}$ asserts that the system
starts in an initial state and every step either satisfies $Next$ or
leaves all variables unchanged (stuttering). An invariant
$Inv$ must hold in every reachable state.

The TLC model checker~\cite{yu1999model} exhaustively explores the
reachable state space for a finite instantiation of the specification
(bounded constants). When TLC finds a state violating an invariant, it
produces a \emph{counterexample}---a minimal execution trace from the
initial state to the violating state. This trace is the key artifact in
our pipeline: Phase~1 extracts it as JSON, and Phase~2 replays it
against a live implementation.

We use per-invariant checking: rather than bundling all 11 invariants
into a single conjunction (which causes TLC to stop at the first
violation), we generate a separate TLC configuration for each invariant.
This produces one counterexample per violated principle, enabling
precise attribution and independent confirmation.

\subsection{Threat Model}\label{sec:threat}

We extend the classical Dolev-Yao adversary~\cite{dolev1983security}
with three LLM-specific threat categories. The Dolev-Yao model assumes
the adversary controls the network: it can intercept, modify, replay,
and inject messages. We additionally assume:

\begin{description}
\item[ADV-1 (Prompt Injection).]
  The adversary injects control directives into tool outputs that the
  LLM interprets as instructions, modifying system behavior.
  This is distinct from traditional injection attacks (SQL, XSS) because
  the ``parser'' is a language model whose behavior is probabilistic,
  not deterministic. The MCP specification acknowledges this risk
  (``servers \textsc{must} sanitize tool outputs'') but the reference
  implementation does not comply.

\item[ADV-2 (Capability Inflation).]
  The adversary forges or inflates capability manifests (MCP tool
  declarations, A2A agent cards) to gain access to tools or operations
  beyond the agent's authorization. This exploits the absence of
  cryptographic attestation on manifests.

\item[ADV-3 (Delegation Amplification).]
  The adversary exploits transitive delegation to re-delegate
  capabilities beyond the scope originally granted. In A2A, an agent
  that receives capability~$c$ via delegation can re-delegate~$c$ to a
  third agent---and the protocol has no mechanism to detect or prevent
  this amplification.
\end{description}

In our \tlaplus{} models, each adversary capability is represented as
an explicit \emph{Adversary action}, separated from Protocol and
Environment actions. This separation ensures that every counterexample
trace clearly identifies whether the violation requires adversarial
participation or arises from normal protocol operation alone.

\subsection{Specification Extraction and Formalization}

Converting protocol documentation into formal models is a well-studied
problem. Pacheco et al.~\cite{pacheco2022} extract protocol-independent
information from RFC prose and compile it into FSMs for automated attack
synthesis. Hermes~\cite{hermes2024} uses a staged pipeline (parser,
transition components, DSL, logical formulas, FSM) to discover
vulnerabilities in cellular network specifications. SAGE~\cite{sage2021}
detects ambiguity and under-specification before code generation.

We follow this tradition with a key adaptation for agent protocols:
our Protocol~IR (\S\ref{sec:agentconform}) explicitly separates
\emph{Protocol}, \emph{Environment}, and \emph{Adversary} actions,
and tags every property with a taxonomy
(\textsc{spec-mandated}/\textsc{spec-recommended}/\textsc{aasm-hardening}/\textsc{aps-completeness})
that distinguishes protocol non-conformance from framework-imposed
security requirements. This prevents the common reviewer objection
that ``the model was designed to find what the authors wanted to find.''

\section{Agent Protocol Stack}\label{sec:aps}

We introduce the \emph{Agent Protocol Stack} (\aps{}), a 6-layer
architectural model that decomposes agent protocol functionality into
distinct security-relevant layers. Each layer specifies what a compliant
protocol \emph{must} address; gaps at any layer propagate upward.

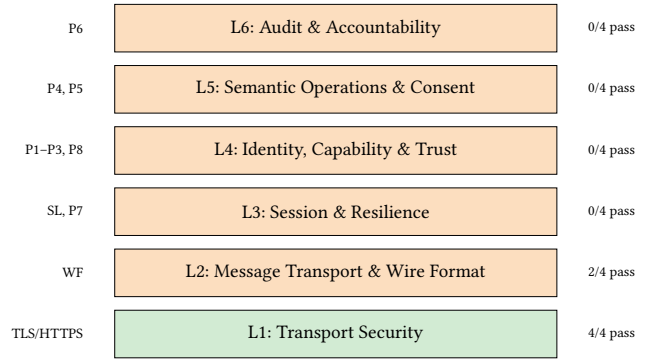
\begin{figure}[t]
\centering
\resizebox{\columnwidth}{!}{%
\begin{tikzpicture}[
  layer/.style={rectangle, draw, minimum width=6.5cm, minimum height=0.7cm,
                align=center, font=\small},
  secure/.style={layer, fill=cellpass!25},
  partial/.style={layer, fill=specfail!25},
  label/.style={font=\scriptsize, anchor=east},
]
\node[secure]  (l1) at (0, 0)   {L1: Transport Security};
\node[partial] (l2) at (0, 0.9) {L2: Message Transport \& Wire Format};
\node[partial] (l3) at (0, 1.8) {L3: Session \& Resilience};
\node[partial] (l4) at (0, 2.7) {L4: Identity, Capability \& Trust};
\node[partial] (l5) at (0, 3.6) {L5: Semantic Operations \& Consent};
\node[partial] (l6) at (0, 4.5) {L6: Audit \& Accountability};

\node[label] at (-3.6, 0)   {TLS/HTTPS};
\node[label] at (-3.6, 0.9) {WF};
\node[label] at (-3.6, 1.8) {SL, P7};
\node[label] at (-3.6, 2.7) {P1--P3, P8};
\node[label] at (-3.6, 3.6) {P4, P5};
\node[label] at (-3.6, 4.5) {P6};

\node[font=\scriptsize, anchor=west] at (3.6, 0)   {4/4 pass};
\node[font=\scriptsize, anchor=west] at (3.6, 0.9) {2/4 pass};
\node[font=\scriptsize, anchor=west] at (3.6, 1.8) {0/4 pass};
\node[font=\scriptsize, anchor=west] at (3.6, 2.7) {0/4 pass};
\node[font=\scriptsize, anchor=west] at (3.6, 3.6) {0/4 pass};
\node[font=\scriptsize, anchor=west] at (3.6, 4.5) {0/4 pass};
\end{tikzpicture}%
}
\caption{The Agent Protocol Stack (\aps{}). Green indicates all protocols
pass; orange indicates at least one protocol has a gap. Left labels show
mapped \aasm{} principles; right labels show pass rates.}
\label{fig:aps-stack}
\end{figure}

\begin{table}[t]
\centering
\caption{APS layers and gap summary across four protocols.
  \CELLPASS{}~=~specified, \SPECFAIL{}~=~spec gap, \UNDERCON{}~=~underconstrained.}
\label{tab:aps-layers}
\small
\begin{tabular}{@{}clcccc@{}}
\toprule
\textbf{Layer} & \textbf{Name} & \textbf{MCP} & \textbf{A2A} & \textbf{ANP} & \textbf{ACP} \\
\midrule
L1 & Transport Security     & \CELLPASS & \CELLPASS & \CELLPASS & \CELLPASS \\
L2 & Message Transport      & \SPECFAIL & \CELLPASS & \UNDERCON & \CELLPASS \\
L3 & Session \& Resilience  & \SPECFAIL & \SPECFAIL & \UNDERCON & \SPECFAIL \\
L4 & Identity \& Trust      & \SPECFAIL & \SPECFAIL & \UNDERCON & \SPECFAIL \\
L5 & Semantic Ops \& Consent& \SPECFAIL & \SPECFAIL & \UNDERCON & \SPECFAIL \\
L6 & Audit \& Accountability& \SPECFAIL & \SPECFAIL & \UNDERCON & \SPECFAIL \\
\bottomrule
\end{tabular}
\end{table}

\paragraph{L1: Transport Security.}
All four protocols inherit TLS/HTTPS from the underlying transport.
This layer is complete across all protocols.

\paragraph{L2: Message Transport \& Wire Format.}
A2A (OpenAPI/gRPC) and ACP (JSON-RPC~2.0) provide complete framing.
MCP's stdio transport lacks message boundary specification, and ANP's
wire format is minimal.

\paragraph{L3: Session \& Resilience.}
All four protocols have incomplete session state machines. MCP sessions
are created implicitly with no explicit close/revoke/suspend semantics.
A2A and ACP have partial lifecycle specifications. ANP's session close
semantics are described as ``vague'' in the white paper.

\paragraph{L4: Identity, Capability \& Trust.}
No protocol provides cryptographic binding of capability manifests.
Credential revocation is unspecified in MCP and ANP; partially
addressed in A2A and ACP.

\paragraph{L5: Semantic Operations \& Consent.}
One protocol allows tool outputs to modify system prompts without
integrity checks. Another models consent as advisory, not enforced.
ACP has emerging consent gates but they are not yet normative.

\paragraph{L6: Audit \& Accountability.}
No protocol mandates comprehensive audit trails. Logging is optional
in MCP, unspecified in ANP, and not normatively required in A2A and ACP.

\paragraph{Observation.}
Every protocol has gaps in at least 4 of the 6 layers. Only L1
(transport security) is universally complete---every other layer
exhibits at least one gap across the four protocols. This motivates
a formal security analysis that goes beyond transport-level properties.

\section{Agent-Agnostic Security Model}\label{sec:aasm}

The \emph{Agent-Agnostic Security Model} (\aasm{}) defines 11 security
principles that any agent protocol should satisfy, independent of its
specific architecture. Each principle is formalized as a \tlaplus{}
invariant that the TLC model checker can verify automatically.

\subsection{Core Security Principles (P1--P8)}

\begin{description}
\item[P1 --- Identity Verifiability.]
  Every agent participating in the protocol must present a verifiable
  identity. Formally: no protocol action succeeds with an unauthenticated
  principal.

\item[P2 --- Capability Attestation.]
  Tool and capability manifests must be cryptographically bound to their
  declaring agent. Forged manifests must be rejected.

\item[P3 --- Delegation Monotonicity.]
  When agent $A$ delegates capabilities to agent $B$, the delegated set
  must be a subset of $A$'s own capabilities:
  $\mathit{delegation}[A][B] \subseteq \mathit{original\_caps}[A]$.
  Transitive re-delegation must not amplify scope.

\item[P4 --- Prompt Integrity.]
  System prompts and control directives must not be modifiable by
  external inputs (tool outputs, agent messages). This formalizes
  protection against prompt injection (ADV-1).

\item[P5 --- Consent Explicitness.]
  Sensitive operations (tool invocation, delegation, data access)
  require explicit user consent before execution, not merely advisory
  notification.

\item[P6 --- Audit Completeness.]
  Every protocol operation must produce a corresponding audit record.
  Formally: $\mathit{audit\_count} \geq \mathit{msg\_count}$ at all
  reachable states.

\item[P7 --- Fail-Secure Defaults.]
  When a protocol encounters an error or ambiguous state, it must
  default to the most restrictive (secure) behavior, not the most
  permissive.

\item[P8 --- Credential \& Registry Integrity.]
  Credentials must be revoked when sessions close. Formally:
  $\mathit{session\_state}[s] = \text{``CLOSED''} \Rightarrow
   \mathit{credentials}[s] = \text{``REVOKED''}$.
\end{description}

\subsection{Completeness Principles (WF, SL, CS)}

\begin{description}
\item[WF --- Wire Format Integrity.]
  All protocol messages must have well-defined structure and valid
  action types.

\item[SL --- Session Lifecycle.]
  Session state machines must cover all reachable states (including
  suspend, resume, and error recovery).

\item[CS --- Composition Safety.]
  Security properties that hold for individual protocols must
  survive when protocols are composed through shared infrastructure
  (bridges, conductors, proxies). This is the subject of
  \S\ref{sec:composition}.
\end{description}

\subsection{Property Taxonomy}\label{sec:aasm:taxonomy}

A critical distinction in our analysis is that \emph{not all invariant
violations represent the same kind of finding}. Protocol specifications
mix RFC~2119 requirement levels (MUST, SHOULD, MAY) with implicit
assumptions and absent requirements. To avoid conflating ``protocol bug''
with ``protocol could be stricter,'' we classify every checked property
into one of four categories:

\begin{description}
\item[\textsc{Spec-Mandated}.]
  Directly required by the protocol specification using MUST-level
  language. A violation is a \emph{standards non-conformance}.
  Example: MCP's Tools specification states ``Servers MUST sanitize tool
  outputs,'' yet the reference SDK performs zero sanitization (P4).

\item[\textsc{Spec-Recommended}.]
  SHOULD-level or best-practice guidance. A violation is a
  \emph{hardening gap}, not a standards violation.
  Example: MCP Authorization states ``Servers SHOULD enforce token
  expiration and rotation'' (P8).

\item[\textsc{AASM-Hardening}.]
  Required by \aasm{} but not addressed by the protocol specification
  at any requirement level. A violation identifies a \emph{design gap}
  where the protocol's specification is silent on a security-relevant
  property.
  Example: A2A has no delegation scope constraints, consent model, or
  audit requirements---all are NOT\_SPECIFIED in the spec (P3, P5, P6).

\item[\textsc{APS-Completeness}.]
  Architectural requirement for a complete protocol stack per the APS
  model (\S\ref{sec:aps}). A violation identifies a \emph{structural
  gap}---an entire protocol layer that is underspecified.
  Example: ANP's wire format is minimal (WF) and session lifecycle is
  vague (SL), both structural gaps at APS layers L2 and L3.
\end{description}

This taxonomy is essential for responsible disclosure. A
\textsc{Spec-Mandated} violation (e.g., P4 in MCP) can be reported as
a standards compliance issue; a \textsc{AASM-Hardening} violation
(e.g., P3 in A2A) is better framed as a feature request or spec
contribution. Table~\ref{tab:taxonomy-examples} shows representative
examples across all four categories.

\begin{table}[t]
\caption{Property taxonomy examples across protocols. SM\,=\,\textsc{Spec-Mandated}, SR\,=\,\textsc{Spec-Recommended}, AH\,=\,\textsc{AASM-Hardening}, AC\,=\,\textsc{APS-Completeness}. Modality abbr.: NS\,=\,\textsc{Not\_Specified}.}\label{tab:taxonomy-examples}
\small
\begin{tabularx}{\columnwidth}{lll>{\raggedright\arraybackslash}X}
\toprule
\textbf{Property} & \textbf{Proto.} & \textbf{Class} & \textbf{Modality} \\
\midrule
P4 Prompt Integrity & MCP & SM & MUST sanitize \\
P7 Fail-Secure & MCP & SM & MUST return 401 \\
P8 Credential Revocation & MCP & SR & SHOULD expire \\
P5 Consent & MCP & SR & SHOULD consent \\
P6 Audit & MCP & SR & SHOULD log \\
P2 Capability Attestation & MCP & AH & NS \\
P3 Delegation & A2A & AH & NS \\
WF Wire Format & ANP & AC & NS \\
SL Session Lifecycle & ANP & AC & NS \\
\bottomrule
\end{tabularx}
\end{table}

\subsection{Counterexample Triage}\label{sec:aasm:triage}

When TLC finds a counterexample, we classify the finding using an
extended taxonomy that accounts for extraction errors and specification
ambiguity:

\begin{description}
\item[\textsc{Spec-Fail}.]
  The protocol allows an execution that violates the property because
  the requirement is absent or too weak in the specification.

\item[\textsc{Impl-Fail}.]
  The protocol specification forbids the behavior, but the reference
  implementation permits it (a compliance bug).

\item[\textsc{Both-Fail}.]
  The specification is weak \emph{and} the implementation is unsafe.

\item[\textsc{Model-Fail}.]
  The formalization introduced an unsupported assumption---the finding
  is an artifact of abstraction, not a real vulnerability.

\item[\textsc{Ambiguity-Fail}.]
  The source documents do not uniquely determine the behavior. The
  counterexample reveals specification ambiguity rather than a definite
  vulnerability.
\end{description}

\textsc{Model-Fail} and \textsc{Ambiguity-Fail} are critical for
methodological honesty: they explicitly acknowledge that formal
verification of informally-specified protocols can produce false
positives. 
ANP accounts for the
highest proportion of \textsc{Ambiguity-Fail} findings due to its
white-paper status.

\subsection{Formalization in TLA\textsuperscript{+}}

Each protocol is modeled as a \tlaplus{} specification with:
\begin{itemize}
\item State variables mirroring protocol state (sessions, credentials,
  capabilities, audit counters).
\item Protocol actions modeling both honest behavior and adversarial
  actions (ADV-1 through ADV-3).
\item \textbf{Protocol actions} modeling honest behavior specified in
  the protocol documentation.
\item \textbf{Environment actions} modeling nondeterministic but
  non-malicious events (credential expiry, transport errors).
\item \textbf{Adversary actions} modeling explicit attacker capabilities
  (ADV-1 prompt injection, ADV-2 capability inflation, ADV-3
  delegation amplification).
\item \aasm{} invariants expressed as state predicates that TLC checks
  at every reachable state.
\end{itemize}


Figure~\ref{fig:p3-invariant} shows the P3 invariant for A2A.

\begin{figure}[t]
\begin{lstlisting}[caption={P3 (Delegation Monotonicity) invariant in the A2A model.},label={fig:p3-invariant},basicstyle=\ttfamily\footnotesize]
P3_DelegationMonotonicity ==
  \A ag1 \in AgentID :
    \A ag2 \in AgentID :
      (ag1 # ag2) =>
        delegation[ag1][ag2]
          \subseteq original_caps[ag1]
\end{lstlisting}
\end{figure}

When TLC finds a state where this predicate is false, it produces a
counterexample trace---a sequence of states from the initial state to
the violating state---that we use as both evidence of the vulnerability
and input for implementation-level testing.

\section{\sys{}: Two-Phase Conformance Checker}\label{sec:agentconform}

\sys{} is a two-phase conformance testing framework that bridges
formal specification analysis and implementation testing through
a provenance-preserving formalization pipeline.

\subsection{Architecture Overview}

\begin{figure}[t]
\centering
\resizebox{\columnwidth}{!}{%
\begin{tikzpicture}[
  box/.style={
    rectangle, draw, rounded corners,
    minimum width=1.3cm,
    minimum height=0.55cm,
    inner sep=2pt,
    align=center,
    font=\scriptsize
  },
  irbox/.style={
    rectangle, draw, rounded corners,
    minimum width=1.3cm,
    minimum height=0.55cm,
    inner sep=2pt,
    align=center,
    font=\scriptsize,
    fill=gray!15
  },
  arrow/.style={-{Stealth[length=1.6mm]}, semithick},
  phase/.style={font=\scriptsize\bfseries, anchor=west}
]

\node[phase] at (-3.6, 2.35) {Formalization};
\node[box]   (spec)   at (-0.9, 2.2) {Spec\\Documents};
\node[irbox] (clause) at (1.2, 2.2)  {Normative\\Clauses};
\node[irbox] (ir)     at (3.3, 2.2)  {Protocol\\IR};
\node[box]   (tla)    at (5.4, 2.2)  {\tlaplus{}\\Model};

\node[phase] at (-3.6, 1.05) {Phase 1: Spec};
\node[box]   (tlc)  at (0.2, 0.9) {TLC Model\\Checker};
\node[box]   (cx)   at (2.8, 0.9) {Counterexample\\Traces};
\node[box]   (tax)  at (5.4, 0.9) {Property\\Taxonomy};

\node[phase] at (-3.6, -0.25) {Phase 2: Impl};
\node[box]   (gen)   at (0.2, -0.4) {Test\\Generator};
\node[box]   (adapt) at (2.8, -0.4) {Protocol\\Adapter};
\node[box]   (result)at (5.4, -0.4) {Conformance\\Report};

\draw[arrow] (spec) -- (clause);
\draw[arrow] (clause) -- (ir);
\draw[arrow] (ir) -- (tla);

\draw[arrow] (tla) -- (tlc);
\draw[arrow] (tlc) -- (cx);
\draw[arrow] (cx) -- (tax);

\draw[arrow] (cx) -- (gen);
\draw[arrow] (gen) -- (adapt);
\draw[arrow] (adapt) -- (result);

\node[font=\tiny, above] at (0.15, 2.2) {extract};
\node[font=\tiny, above] at (2.25, 2.2) {compile};
\node[font=\tiny, above] at (4.35, 2.2) {generate};
\node[font=\tiny, above] at (1.5, 0.9) {\aasm{}};
\node[font=\tiny, above] at (4.1, 0.9) {classify};

\end{tikzpicture}%
}
\caption{\sys{} architecture. Spec documents are formalized through a typed Protocol~IR before generating \tlaplus{} models. Phase~1 finds spec-level violations; Phase~2 replays traces against live implementations. Shaded boxes indicate the IR layer.}
\label{fig:architecture}
\vspace{-0.5em}
\end{figure}
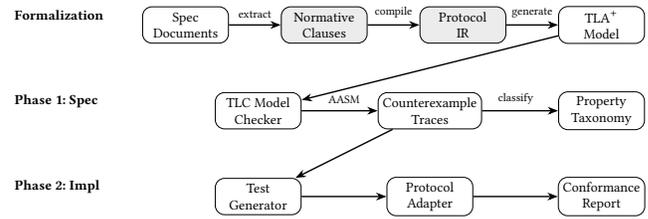

The key design principle is that \emph{no \tlaplus{} action or invariant
is written directly from informal intuition}. Every formal artifact is
traceable to either (a)~an extracted protocol clause,
(b)~an explicit environment assumption,
(c)~an explicit attacker capability, or
(d)~an \aasm{} principle.

\subsection{Formalization Pipeline}

\paragraph{Step 1: Source collection.}
For each protocol we collect all authoritative specification artifacts:
normative prose, schemas, examples, diagrams, and reference
implementations. Sources are assigned a precedence order
(normative text $>$ schemas $>$ diagrams $>$ implementation $>$
conventions). Conflicts are recorded as ambiguities rather than
silently resolved.

\paragraph{Step 2: Normative clause extraction.}
We extract every clause that affects behavior, safety, authority,
lifecycle, or trust boundaries. Each clause is stored with its
modality (\textsc{must}, \textsc{should}, \textsc{may}, or
\textsc{not\_specified} per RFC~2119~\cite{rfc2119}), source reference,
actor binding, and an explicit ambiguity flag. For MCP, we extracted
37~normative clauses from 8~specification documents.

\paragraph{Step 3: Protocol IR construction.}
Following the staged pipeline approach of Pacheco et
al.~\cite{pacheco2022}, Hermes~\cite{hermes2024}, and
PROSPER~\cite{prosper2023}, we convert
extracted clauses into a typed intermediate representation rather than
mapping directly into \tlaplus{}. Each IR transition record carries:
\texttt{actor}, \texttt{trigger}, \texttt{preconditions},
\texttt{state\_writes}, \texttt{modality}, \texttt{kind}~$\in$~\{\textit{Protocol},
\textit{Environment}, \textit{Adversary}\}, and
\texttt{source \_refs}.
Each IR property record carries a class tag from:

\begin{description}
\item[\textsc{spec-mandated}:] Directly required by the protocol (\textsc{must}).
\item[\textsc{spec-recommended}:] \textsc{should}-level requirement.
\item[\textsc{aasm-hardening}:] Required by \aasm{} but not by the protocol.
\item[\textsc{aps-completeness}:] Architectural requirement from \aps{}.
\end{description}

This taxonomy is critical because it separates ``the protocol has a
bug'' (\textsc{spec-mandated} violation) from ``the protocol could be
more secure'' (\textsc{aasm-hardening} gap).

\paragraph{Step 4: IR to \tlaplus{} compilation.}
We compile the IR into \tlaplus{} in a syntax-directed way:
actors become quantified principals, triggers become guarded actions,
state reads/writes become primed variable updates.
\textsc{must}~requirements become safety invariants;
\textsc{should}~requirements become hardening checks;
\textsc{not\_specified} items become underspecification findings.

The three action kinds are kept separate in the model:
\emph{Protocol actions} are transitions licensed by the spec;
\emph{Environment actions} model nondeterministic but non-malicious
events (e.g., token expiry); \emph{Adversary actions} model explicit
attacker capabilities from our Dolev-Yao+LLM threat model
(\S\ref{sec:threat}). This prevents the reviewer objection that ``the
attacker was inserted by the modeler.''

\subsection{Phase 1: Spec-Level Analysis}

Phase~1 checks each \aasm{} invariant independently using TLC.

\paragraph{Per-invariant checking.}
Rather than bundling all invariants into a single conjunction (which
causes TLC to stop at the first violation), we generate a separate TLC
configuration for each invariant. This produces one counterexample
per violated principle, enabling precise attribution.

\paragraph{Counterexample extraction and triage.}
The \texttt{tlc\_runner.py} tool parses TLC output to extract:
(i)~which invariant was violated,
(ii)~the depth of the counterexample,
(iii)~the full state trace with variable values at each step.
Each counterexample is classified as:

\begin{description}
\item[\textsc{spec-fail}:] Protocol allows the violation (requirement absent/weak).
\item[\textsc{impl-fail}:] Protocol forbids it, but implementation permits it.
\item[\textsc{both-fail}:] Protocol is weak and implementation is also unsafe.
\item[\textsc{model-fail}:] Formalization introduced an unsupported assumption.
\item[\textsc{ambiguity-fail}:] Source documents do not uniquely determine behavior.
\item[\textsc{pass}:] Property holds at both levels.
\end{description}

\textsc{model-fail} and \textsc{ambiguity-fail} are essential for
methodological honesty: they provide explicit buckets for extraction
errors and underspecified source text, which prior work on
spec-to-formal conversion has identified as first-class
outcomes~\cite{sage2021}.

\subsection{Phase 2: Implementation-Level Testing}

Phase~2 converts TLC counterexample traces into executable test cases
that run against live protocol implementations. The same IR property
that generated the invariant also generates the implementation test
oracle.

\paragraph{Protocol adapters.}
Each protocol has an adapter class (subclass of
\texttt{ProtocolAdapter}) that maps \tlaplus{} action names to concrete
SDK or API calls:

\begin{itemize}
\item \textbf{MCP adapter:} 7~protocol actions $\to$ JSON-RPC calls
  (\texttt{ini-} \texttt{tialize}, \texttt{tools/list},
  \texttt{tools/call}, \texttt{shutdown}).
\item \textbf{A2A adapter:} 6~protocol actions $\to$ HTTP calls
  (agent discovery, task send, delegation, consent).
\end{itemize}

\paragraph{Replay and invariant checking.}
The \texttt{replay\_counterexample()} method executes each action in
the trace sequence, then checks the relevant invariant at the
implementation level. We run live tests against reference SDK
implementations of two protocols with mature Python SDKs, plus
official reference servers, for a total of 42 tests. This
counterexample-to-conformance workflow follows the tradition of
rigorous protocol conformance testing~\cite{bishop2005rigorous}, while
adapting the oracle source to formally checked \tlaplus{} traces.

\paragraph{Mock vs.\ live modes.}
Adapters support both mock mode (offline, deterministic) and live mode
(against real servers). Mock mode enables reproducible CI testing;
live mode confirms real-world impact. The live test suite for MCP
includes source code analysis (checking for absent sanitization,
missing hooks) and behavioral tests (injecting payloads, calling
undeclared tools, closing sessions without credential revocation).

\section{Composition Safety}\label{sec:composition}

Our most significant conceptual contribution is the \emph{Composition
Safety} (CS) principle: security properties that hold for individual
protocols can break when protocols are composed through shared
infrastructure. This section presents the principle, the formal
framework for analyzing it, and preliminary findings.

\subsection{Motivation}

Agent protocols are increasingly deployed together. A conductor or
proxy may bridge a tool-invocation protocol with a delegation protocol,
routing tool outputs from one protocol into delegation decisions in
another. Chained deployments are common: a client connects to a
server that internally proxies requests to another server. In
multi-protocol and chained deployments, the \emph{composition
boundary}---the bridge, conductor, or proxy connecting two protocol
domains---becomes an ungoverned attack surface.

Three composition patterns are particularly concerning:

\begin{enumerate}
\item \textbf{Cross-protocol cascade.}
  A design gap in Protocol~A (e.g., no output sanitization)
  propagates through a bridge to cause a security failure in
  Protocol~B (e.g., delegation amplification). Neither protocol's
  individual analysis detects this.

\item \textbf{Chained server injection.}
  Server~A proxies tool calls to Server~B. Content injected at
  Server~B passes through Server~A to the client without sanitization.
  The client implicitly trusts Server~B through Server~A, with no
  mechanism to verify the trust chain.

\item \textbf{Consent bypass via protocol routing.}
  A sensitive operation that requires consent in Protocol~A can be
  routed through Protocol~B (which has no consent mechanism),
  bypassing the consent gate entirely.
\end{enumerate}

\subsection{Formal Framework}

We model composition by combining two protocol models with a shared
bridge component. The bridge has state variables that capture
cross-protocol contamination:

\begin{itemize}
\item $\mathit{bridge\_compromised}$: whether a design gap in one
  protocol has tainted the bridge.
\item $\mathit{credential\_forwarded}$: whether credentials from one
  protocol have leaked into another.
\item Operation counters that track whether bridge actions produce
  audit records in either protocol's audit domain.
\end{itemize}

CS invariants check whether security properties survive composition:

\begin{description}
\item[CS\_NoLeakage:]
  A design gap in Protocol~A must not affect Protocol~B's
  capability integrity. Formally: if the bridge is compromised by
  Protocol~A, no agent in Protocol~B gains capabilities beyond its
  original grant.

\item[CS\_IsolationHolds:]
  The presence of a design gap in one protocol must not correlate
  with state changes in the other.

\item[CS\_AuditChain:]
  Bridge operations must be visible in at least one protocol's
  audit trail.
\end{description}

\subsection{Composition Patterns Analyzed}

We construct five composed \tlaplus{} models covering the major
protocol composition patterns:

\begin{enumerate}
\item \textbf{Tool protocol + Delegation protocol:} tool output
  injection cascades through bridge to delegation amplification.
\item \textbf{Chained tool servers:} injection chain, credential
  forwarding, transitive trust collapse, manifest inflation.
\item \textbf{Tool protocol + Capability protocol:} consent bypass
  via routing through the protocol without consent mechanisms.
\item \textbf{Delegation + Capability protocol:} authority model
  conflict and credential lifecycle mismatch.
\item \textbf{Federated delegation:} delegation monotonicity breaks
  at federation boundaries due to absent mutual verification.
\end{enumerate}

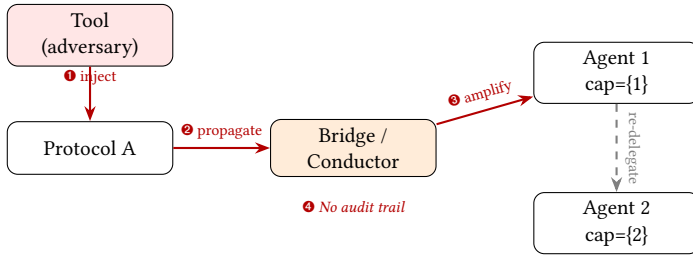
\begin{figure}[t]
\centering
\begin{tikzpicture}[
  box/.style={rectangle, draw, rounded corners, minimum width=2.2cm,
              minimum height=0.7cm, align=center, font=\small},
  attack/.style={-{Stealth[length=2mm]}, thick, red!70!black},
  normal/.style={-{Stealth[length=2mm]}, thick},
  label/.style={font=\scriptsize, midway, above},
]
\node[box, fill=red!10] (tool) at (0, 0) {Tool\\(adversary)};
\node[box] (protoA) at (0, -1.5) {Protocol A};

\node[box, fill=orange!15] (bridge) at (3.5, -1.5) {Bridge /\\Conductor};

\node[box] (ag1) at (7, -0.5) {Agent 1\\cap=\{1\}};
\node[box] (ag2) at (7, -2.5) {Agent 2\\cap=\{2\}};

\draw[attack] (tool) -- node[label] {\ding{202} inject} (protoA);
\draw[attack] (protoA) -- node[label] {\ding{203} propagate} (bridge);
\draw[attack] (bridge) -- node[label, sloped] {\ding{204} amplify} (ag1);
\draw[normal, dashed, gray] (ag1) -- node[label, sloped] {\scriptsize re-delegate} (ag2);

\node[font=\scriptsize\itshape, text=red!70!black, below=0.2cm of bridge]
  {\ding{205} No audit trail};
\end{tikzpicture}
\caption{Abstract cross-protocol attack chain: a design gap in
Protocol~A propagates through a shared bridge to amplify delegation
in Protocol~B. Bridge operations fall outside both protocols' audit
domains.}
\label{fig:composition-attack}
\end{figure}

\subsection{Preliminary Findings}

Across 5~composed models with 21~CS invariants, TLC finds violations
in 20---only one invariant (domain isolation in the federated model)
holds by construction. The key findings are:

\begin{enumerate}
\item \textbf{Individual protocol analysis is necessary but
  insufficient.} None of the 20 composition violations are detectable
  by analyzing protocols individually.

\item \textbf{Chained server composition is the highest-risk pattern.}
  All five invariants are violated in the chained-server model,
  including credential forwarding and transitive trust collapse---in
  the most widely deployed composition pattern.

\item \textbf{Consent bypass via protocol routing is a confused deputy.}
  Routing operations through a protocol with weaker security bypasses
  the consent gates of the stronger protocol.

\item \textbf{Bridge components are ungoverned.}
  No protocol specifies security properties for bridge operations.
  Bridges fall outside both protocols' audit domains.
\end{enumerate}

Full counterexample traces and implementation-level confirmation will
be presented in the complete version after disclosure timelines
conclude.

\section{Related Work}\label{sec:related}

\paragraph{Protocol security frameworks.}
Internet protocol engineering relies on security considerations
frameworks (e.g., RFC~3552~\cite{rfc3552}) to enforce threat-model
and trust-boundary analysis during design. Our \aps{} and \aasm{}
serve an analogous role for agent protocols, where semantic payloads
and delegation chains introduce failure modes beyond traditional
message-security analysis.

\paragraph{Formal methods for protocols.}
Formal protocol verification has a rich history: CSP-based
attacks~\cite{lowe1996breaking},
ProVerif~\cite{blanchet2001proverif},
Tamarin~\cite{meier2013tamarin}, and surveys of formal methods for
deployed protocols~\cite{basin2018formal, bhargavan2002formal}.
We differ in targeting agent protocols with LLM-specific threats,
using \tlaplus{}/TLC for bounded model checking with direct
counterexample extraction, and introducing composition safety as a
first-class property.

\paragraph{LLM and agent security.}
Indirect prompt injection~\cite{greshake2023youve}, tool-use
vulnerabilities~\cite{zhan2024injecagent}, and agent security
benchmarks~\cite{debenedetti2024agentdojo} emphasize attack discovery.
Our work is complementary: we target specification-level conformance
with traceable links from protocol clauses to formal properties.

\paragraph{Protocol composition.}
Cortier and Delaune~\cite{cortier2009safely} show that security
properties may not compose even when individual protocols are secure.
Our CS principle adapts this insight to agent protocols, where
composition occurs through semantic bridges rather than cryptographic
primitives.

\paragraph{Specification extraction.}
Pacheco et al.~\cite{pacheco2022} extract protocol information from
RFC prose into FSMs. Hermes~\cite{hermes2024} uses a staged pipeline
for cellular protocols. SAGE~\cite{sage2021} detects specification
ambiguity. Li et al.~\cite{li2025extracting} separate annotation from
formalization. ParCleanse~\cite{zheng2025parcleanse} maintains
RFC-to-test traceability. Basin et al.~\cite{basin2025bridging} argue
that complete protocol understanding requires multiple artifact types.
Our pipeline follows this tradition with a typed Protocol~IR, explicit
action separation, and a property taxonomy for agent-specific threats.

\paragraph{Agent protocol standards.}
MCP~\cite{mcp2024}, A2A~\cite{a2a2024}, ANP~\cite{anp2025},
ACP~\cite{acp2025}, and ACP-Client~\cite{acpclient2025} represent
the emerging agent protocol ecosystem. To our knowledge, no prior
work provides a systematic formal security analysis across multiple
agent protocols.

\section{Discussion}\label{sec:discussion}

\subsection{Responsible Disclosure}\label{sec:discussion:disclosure}

We have begun applying \sys{} to live SDK implementations and
official reference servers. Implementation-level testing is ongoing;
this preprint focuses on the framework and methodology rather than
specific findings. Confirmed findings will be reported through
coordinated disclosure before publication of the full evaluation.

\subsection{Limitations}

\paragraph{Bounded model checking.}
TLC explores a finite state space bounded by constants. Our models
use small bounds (2--3 agents, 2 capabilities). Counterexamples
found within these bounds are valid, but the absence of a violation
does not guarantee correctness for larger instances.

\paragraph{Specification fidelity.}
The IR extraction pipeline depends on human interpretation of
protocol documentation. While the IR provides traceability, the
extraction itself is not fully automated and may introduce
\textsc{model-fail} errors. Our taxonomy explicitly accounts for this.

\paragraph{Protocol drift.}
Agent protocols evolve rapidly. Results correspond to specification
snapshots and should be interpreted as versioned evidence.

\paragraph{Adversary model scope.}
Our adversary model (Dolev-Yao + ADV-1/2/3) targets protocol-level
threats. Side-channel attacks, timing attacks, and physical access
are out of scope.


\bibliographystyle{ACM-Reference-Format}
\bibliography{references}


\end{document}